\documentclass{article}

\usepackage{amsmath} 

\numberwithin{equation}{section}

\usepackage{amsfonts}

\usepackage{slashed}

\def\be{\begin{equation}}
\def\ee{\end{equation}}
\def\bea{\begin{eqnarray}}
\def\eea{\end{eqnarray}}

\def\({\left(}
\def\){\right)}
\def\<{\left<}
\def\>{\right>}

\def\tr{{\mbox{tr}}}
\def\be{\begin{equation}}
\def\ee{\end{equation}}
\def\bea{\begin{eqnarray*}}
\def\eea{\end{eqnarray*}}
\def\ben{\begin{eqnarray}}
\def\een{\end{eqnarray}}
\def\({\left(}
\def\){\right)}
\def\<{\left<}
\def\>{\right>}
\def\!{\right|}
\def\|{\left|}

\def\[{\left[}
\def\]{\right]}

\def\+{\bar}
\def\mb{\mathbb}
\def\tr{{\mbox{tr}}}

\def\L{{\cal{L}}}
\def\t{\widetilde}

\def\N{{\cal{N}}}

\def\O{{\cal{O}}}

\def\Z{{\cal{Z}}}

\def\l{\ell}
\def\h{\widehat}

\begin{document}
\setlength{\unitlength}{1mm}

\pagestyle{empty}
\vskip-10pt
\vskip-10pt
\hfill 
\begin{center}
\vskip 3truecm
{\Large \bf
Deconstructing graviphoton}\\
\vskip0.3cm
{\Large \bf from}\\
\vskip0.3cm
{\Large \bf mass-deformed ABJM}\\
\vskip 2truecm
{\large \bf
Andreas Gustavsson\footnote{a.r.gustavsson@swipnet.se}}\\
\vskip 1truecm
{\it  Physics Department, University of Seoul, 13 Siripdae, Seoul 130-743 Korea}
\end{center}
\vskip 2truecm
{\abstract{Mass-deformed ABJM theory has a maximally supersymmetric fuzzy two-sphere vacuum solution where the scalar fields are proportional to the TGRVV matrices. We construct these matrices using Schwinger oscillators. This shows that the ABJM gauge group that corresponds to the fuzzy two-sphere geometry is $U(N)\times U(N-1)$. We deconstruct the graviphoton term in the D4 brane theory. The normalization of this term is fixed by topological reasons. This gives us the correct normalization of the deconstructed $U(1)$ gauge field and fixes the Yang-Mills coupling constant to the value which corresponds to M5 brane compactified on $\mb{R}^{1,2} \times S^3/{\mb{Z}_k}$. The graviphoton term also enable us to show that the zero mode contributions to the partition functions for the D4 and the M5 brane agree.}}

\vfill 
\vskip4pt
\eject
\pagestyle{plain}

\section{Introduction}
Mass-deformed ABJM theory has been constructed and studied in for example \cite{Gomis:2008vc}, \cite{Lee:2009mm}, \cite{Lambert:2009qw}, \cite{Kim:2010mr}. The mass-deformed theory not only preserves maximal supersymmtry, but it also has a maximally supersymmetric fuzzy two-sphere vacuum solution in which the scalar fields are proportional to the TGRVV matrices \cite{Terashima:2008sy}, \cite{Gomis:2008vc}.

Small fluctuations about the fuzzy two-sphere and in a sector with vanishing magnetic flux should be described by the D4 brane action. One expects that this deconstructed theory corresponds to dimensional reduction of an M5 brane on $\mb{R}^{1,2} \times S^3/{\mb{Z}_K}$ along the Hopf fiber by taking $K$ large \cite{Nastase:2009ny}, \cite{Gustavsson:2011af}. One motivation for studying this dimensional reduction is to better understand the relation between D4 and M5 brane worldvolume theories \cite{Lambert:2010iw}, \cite{Douglas:2010iu}. As we have a non-trivial fiber bundle, dimensional reduction should give rise to a graviphoton term in the D4 brane action \cite{Linander:2011jy}, \cite{Witten:2009at}. But the graviphoton term was absent in \cite{Nastase:2009ny}. In this paper we argue that the origin for this, is the the wrong choice of ABJM gauge group as $U(N)\times U(N)$. A problem arises when we want to associate all $N\times N$ matrices with the geometry of a bifundamental fuzzy two-sphere. Namely, this is impossible to do. An alternating product of TGRVV matrices and their hermitian conjugates, only gives us $(N-1)N$ independent matrices \cite{Nastase:2009ny}. Since we are interested in large-$N$ limit one may think that the difference between $(N-1)N$ and $N^2$ would contribute with only subleading terms and so be $\frac{1}{N}$-suppressed. But that is not quite correct. The difference gives rise to one new leading term which is the missing graviphoton term as we will see. 

There is also a difficult, or perhaps impossible to solve, non-commutative geometry problem, concerning the difference. Namely, if we start with gauge group $U(N)\times U(N)$, we have to add by hand $N$ further bi-fundamental matrices, as was done in \cite{Nastase:2009ny}. But these further $N$ matrices can not be mapped into spherical harmonics of a bifundamental fuzzy two-sphere.

If we want to describe a bifundamental two-sphere, then the gauge group must be $U(N) \times U(N-1)$. We may then also deconstruct the graviphoton term from mass-deformed ABJM theory. 

In section \ref{2} we describe the mass-deformed ABJM Lagrangian and consider the vacuum equation of the bifundamental fuzzy two-sphere. In section \ref{3} we describe a Schwinger construction of the TGRVV matrices which clarifies why they are really $N\times (N-1)$ matrices, and not $N\times N$ matrices (though of course we can add a column of zeroes, whose geometrical meaning is then unclear). We also derive the associated gauge group from the three-algebra that is generated by the TGRVV operators and make it plausible that this will be $U(N)\times U(N-1)$ (though we only present the computation in the case of $N=2$). In section \ref{4} we revisit the novel Higgs mechanism and derive the graviphoton and the YM coupling constant which matches with the M5 brane coupling constant which is fixed by selfduality. Indeed this is very remarkable since we start from ABJM so somehow ABJM knows about selfduality and the M5 brane. In section \ref{5} we show how the graviphoton term helps to match zero mode contributions to the partition functions in D4 brane with the M5 brane on $\mb{R}\times T^2 \times S^3/{\mb{Z}_K}$.

\section{The mass-deformed ABJM Lagrangian}\label{2}
We may formulate ABJM theory in a manifestly $SU(4)$ R symmetric way, and we may also use a three-algebra formulation \cite{Bagger:2008se}, \cite{Gustavsson:2010yr}. If the generators are denoted $T^a$, their conjugates are denoted $T_a$, then we may define the three-bracket as
\bea
[T^a,T^b;T^c] &=& T^a T_c T^b - T^b T_c T^a
\eea
and this will satisfy the hermitian fundamental identity
\bea
[[T^a,T^b;T^c],T^d;T^e] &=& [[T^a,T^d;T^e],T^b;T^c] + [T^a,[T^b,T^d;T^e];T^c]\cr
&& - [T^a,T^b;[T^c,T^e;T^d]]
\eea
We have a three-algebra if we can express the three-bracket as a linear combination of generators
\bea
[T^a,T^b;T^c] &=& f^{ab}{}_{cd} T^d
\eea
where index $c$ shall sit down-stairs due to the definition of the three-bracket above. The ABJM Lagrangian is completely fixed once we know the structure constants $f^{ab}{}_{cd}$ of the three-algebra. So we do not really need a matrix realization of the three-algebra generators $T^a$. In this three-algebra language, the bosonic part of the mass-deformed ABJM Lagrangian is given by
\bea
\L &=& \L_{kin} + \L_{pot} + \L_{CS}
\eea
where 
\bea
\L_{kin} &=& -D_{\mu} Z^A_a D^{\mu} Z_A^a\cr
\L_{pot} &=& - W^{AB}_{Ca} W_{AB}^{Ca}\cr
\L_{CS} &=& \frac{1}{2} \epsilon^{\mu\nu\lambda} \(f^{ab}{}_{cd} B_{\mu}{}^c{}_b \partial_{\nu} B_{\lambda}{}^d{}_a + \frac{2}{3} f^{ac}{}_{dg} f^{ge}{}_{fb} B_{\mu}{}^b{}_a B_{\nu}{}^d{}_c B_{\lambda}{}^f{}_e\)\cr
&&+\sum_i \frac{k_i e_i^2}{4\pi} A^i dA^i
\eea
and where 
\bea
D_{\mu} X_a &=& \partial_{\mu} X_a - \t A_{\mu}{}^b{}_a X_b
\eea
and
\bea
W^{AB}_{Ca} &=& \(\delta^{[A}_C Z^{B]}_b Z^C_c Z_C^d + Z^A_b Z^B_c Z_C^d\) f^{bc}{}_{da} + m G^{[A}_C Z^{B]}_a
\eea
The matrix $G^A_B$ responsible for the mass-deformation, shall satisfy $(G^A_B)^* = G^B_A$, $G^A_B G^B_C = \delta^A_C$ and $G^A_A = 0$. There is no such solution which preserves the full $SU(4)$ R symmetry. The best we can do is to break it down to $SU(2) \times SU(2) \times U(1)$, and we may accordingly split the R-symmetry index as $Z^A = (Z^a,Z_{\dot{a}})$ and then the matrix $G^A_B$ has non-vanishing components $G^a_b = \delta^a_b$ and $G_{\dot{a}}^{\dot{b}} = -\delta_{\dot{a}}^{\dot{b}}$. 

Following \cite{Gustavsson:2010yr}, we decompose the three-algebra structure constants as
\bea
f^{bc}{}_{da} &=& \t f^{bc}{}_{da} + \lambda \delta^b_a \delta^c_d
\eea
where
\bea
\t f^{bc}{}_{ca} &=& 0\cr
\t f^{bc}{}_{cb} &=& 0
\eea
We decompose the gauge field as
\bea
\t A^b{}_a &=& \t B^b{}_a + e_i A^i_{\mu} \delta^b_a\cr
\t B^b{}_a &=& B^d{}_c \t f^{bc}{}_{da}
\eea
but if we assume that $B^b{}_b = 0$ then we may just as well write $\t B^b{}_a = B^d{}_c  f^{bc}{}_{da}$.

The three-algebra structure constants are given by
\bea
\t f^{bc}{}_{da} &=& - \sum_{l=1}^L \frac{2\pi}{K_l} g^{A_l B_l} (t_{A_l})^b{}_a (t_{B_l})^c{}_a + \lambda \delta^b_a \delta^c_d
\eea
where $\lambda$ has to be chosen such that $f^{bc}{}_{da} = -f^{cb}{}_{da}$

The CS term now becomes
\bea
\L_{CS} &=& \sum_l \frac{K_l}{4\pi} tr_l \( AdA + \frac{2i}{3} A^3\) + \sum_i \frac{k_i e_i^2}{4\pi} A^i dA^i
\eea
and the covariant derivative becomes
\bea
D_{\mu} Z_a &=& \partial_{\mu} Z_a - i A^A (t_A)^b{}_a Z_b - i e_i A^i_{\mu} Z_a
\eea
Supersymmetry variations are given by
\bea
\delta Z^A_a &=& i \bar{\epsilon}^{AB} \psi_{Ba}\cr
\delta \psi_A &=& -\gamma^{\mu} \epsilon_{AB} D_{\mu} Z^A + \epsilon_{BC} W^{BC}_A\cr
\delta \t A_{\mu}{}^b{}_a &=& -i \bar{\epsilon}_{AB} \gamma_{\mu} Z^A_c \psi^{Bd} f^{bc}{}_{da} + i \bar{\epsilon}^{AB} \gamma_{\mu} \psi_{Ac} Z_B^d f^{bc}{}_{da}
\eea
if we use the convention 
\bea
(\epsilon_{AB})^* &=& \epsilon^{AB}
\eea
These are exactly the conventions used in \cite{Bagger:2008se}. There is nothing wrong with these supersymmetry variations. They close on-shell. However, they are not complete in the sense that they are not enough to verify invariance of the action. For this we also need to obtain supersymmetry variations of $A^i_{\mu}$ such that they leave the action invariant. This can be straightforwardly done although we will not present these supersymmetry variations here. 

The structure constants are now proportional to $\frac{2\pi}{K}$, and later on we will also extract out a factor of $\hbar$ thus making structure constants proportional to $\frac{2\pi \hbar}{K}$. But for deconstruction it will be more convenient to have $\frac{K}{2\pi \hbar}$ as an overall factor multiplying the whole action, and having structure constants $\sim 1$. To this end we rescale the fields as
\bea
Z^A &=& \sqrt{\frac{K}{2\pi \hbar}} {Z'}^A\cr
\psi_A &=& \sqrt{\frac{K}{2\pi \hbar}} \psi'_A\cr
A_{\mu} &=& A'_{\mu}
\eea
After this rescaling, the Lagrangian becomes
\bea
\L &=& \frac{K}{2\pi\hbar} \L'
\eea
where
\bea
\L' &=& \L'_{CS} + \L'_{kin} + L'_{pot}
\eea
and 
\bea
\L'_{CS} &=& \frac{\hbar}{2}\tr_L \(A^L dA^L + \frac{2i}{3}(A^L)^3\) - \frac{\hbar}{2}\tr_R \(A^R dA^R + \frac{2i}{3} (A^R)^3\)\cr
\L'_{kin} &=& -\tr_L \(D_{\mu} Z^A D^{\mu} Z_A\)\cr
\L'_{pot} &=& -\tr_L\(W^{BC}_A W_{BC}^A\)
\eea
where
\bea
W^{AB}_C &=& \frac{1}{\hbar}\(\delta^{[A}_C [Z^{B]},Z^C;Z^C] + [Z^A,Z^B,Z^C]\) + m' G^{[A}_C Z^{B]}
\eea
and 
\bea
m' &=& m
\eea

\subsection{Maximally supersymmetric two-sphere}
The scalar field equations of motion to be satisfied for the static vacuum expectation value are given by
\bea
W^{AB}_C &=& 0
\eea
These are the BPS equations we obtain by requiring $\delta \psi_A = 0$. Since it does not constrain the supersymmetry parameters, any solution to this set of equations will be maximally supersymmetric. In this paper we will take the vacuum expectation value to be on the form
\bea
Z^A_{vev} = v^A = \(\begin{array}{c}
v^a \\
v_{\dot{a}}
\end{array}\)
\eea
and limit ourselves to the following BPS equations
\bea
[v^a,v^b;v^c] &=& -2m \hbar \delta^{ab}_{cd} v^d\cr
v_{\dot{a}} &=& 0
\eea
This means that  
\bea
v^a &=& \sqrt{m\hbar} G^a
\eea
where $G^a$ satisfy the TGRVV three-algebra
\ben
[G^a,G^b;G^c] &=& -2 \delta^{ab}_{cd} G^d\label{TGRVV}
\een
We want to relate TGRVV generators with functions $z^a(\theta,\varphi)$ on $S^2$ on which we denote the spherical coordinates by $\sigma^m = (\theta,\varphi)$. We assume that
\bea
z^a z_a &=& \frac{R^2}{2}
\eea
and we relate these with four real euclidean coordinates as
\bea
z^1 &=& \frac{1}{\sqrt{2}} \(x^1 + i x^2\)\cr
z^2 &=& \frac{1}{\sqrt{2}} \(x^2 + i x^2\)
\eea
We have the three-sphere constraint
\bea
x^i x^i &=& R^2
\eea
and we have
\bea
\{x^i,x^j,x^k\} &=& \frac{1}{R} \epsilon^{ijkl} x^l
\eea
where the Nambu bracket is defined as
\bea
\{x^i,x^j,x^k\} &=& *_{S^3} \(dx^i \wedge dx^j \wedge dx^k\)
\eea
where $*_{S^3}$ is Hodge dual with respect to the induced metric on $S^3$. We can map this to complex coordinates and we find
\bea
\{z^a,z^b,z_c\} &=& \frac{2i}{R} \delta^{ab}_{cd} z^d
\eea

We make the following ansatz for the isomorphism
\ben
G^a &\cong & \lambda z^a\label{z}
\een
Let us assume a unit normalized trace form which we will denote as $\int_{S^2}$ on the space of functions on $S^2$,
\bea
\int_{S^2} 1 &=& 1
\eea
If we assume that the radius on $S^2$ is $\frac{R}{2}$, then we have
\bea
\int_{S^2} &:=& \frac{1}{\pi R^2} \int_0^{\pi} d\theta \int_{0}^{2\pi} d\varphi \frac{R^2}{4} \sin \theta
\eea
Then we have the correspondence (the precise meaning of the trace $\tr_N$ will be explained below, and especially we here use Eq (\ref{radii}) below)
\bea
\frac{1}{N} \tr_N (G^a G_a) &=& \lambda^2 \int_{S^2} z^a z_a
\eea
which determines 
\bea
\lambda &=& \frac{\sqrt{2(N-1)}}{R}
\eea
We define\footnote{The minus sign can be removed here if we map $m \rightarrow -m$ but that will change the TGRVV algebra by a minus sign. We keep our signs this way since it was this way the TGRVV matrices were originally defined. We choose to have a minus sign here in order to have $\hbar$ positive.}${}^{,}$\footnote{In general there is no way to add $\O(\hbar^2)$ terms to the Nambu bracket such that we preserve the hermitian fundamental identity, but here we can do this by simply splitting the Nambu bracket into a sum of Poisson brackets by extracting the phase factor as $z^a = e^{\i\psi} \t z^a$.}
\bea
[z^a,z^b;z^c] &=& -i\hbar \{z^a,z^b,z_c\} + \O(\hbar^2)
\eea
and we get
\bea
[z^a,z^b;z^c] &=& \frac{2 i \hbar}{R} \delta^{ab}_{cd} z^d
\eea
This equals the TGRVV algebra if we take
\bea
\hbar &=& \frac{R^3}{2(N-1)}
\eea
and make the appropriate scalings of $z^a$ by $\lambda$'s to get $G^a$'s.

We thus have
\bea
z^a &=& \frac{R}{\sqrt{2(N-1)}} G^a
\eea
and we have
\bea
v^a &=& \sqrt{m\hbar} G^a
\eea
Identifying these, we conclude that
\bea
m &=& \frac{1}{R}
\eea 
Up to a convention dependent sign, this is the same relation between $m$ and $R$ as we have in mass-deformed Nambu-BLG theory defined on $S^3$ \cite{Gustavsson:2011af}. In this reference the geometrical meaning of the $S^3$ is manifest as the world-volume of the M5 brane. Nambu-BLG theory can also be reformulated as an ABJM theory with mass parameter $m$ by using $SO(8)$ triality.

\section{Schwinger construction of TGRVV algebra}\label{3}
Intuitively we like to think of the TGRVV algebra (\ref{TGRVV}) as corresponding to a non-commutative three-sphere embedded in $\mb{C}^2$. On a technical level what we achieve is a non-commutative two-sphere corresponding to the base-manifold. In any case, at least intuitively it is tempting to make $\mb{C}^2$ non-commutative. Thus if we interpret $G^a$ as corresponding to the two complex coordinates $z^a\in \mb{C}^2$ (up to a rescaling by a factor of $\lambda$ as in Eq (\ref{z})) then we may turn this into a non-commutative space by assuming that
\ben
[G_a,G^b] &=& \delta_a^b\cr
[G^a,G^b] &=& 0\cr
[G_a,G_b] &=& 0\label{osc1}
\een
These commutation relations correspond to non-commutative $\mb{R}^4$ with a self-dual non-commutativity parameter $\lambda^{-2}$. As we will see, these commutation relations imply that the $G^a$ obey the TGRVV algebra (\ref{TGRVV}). We note that $G^a$ now become creation operators, and the conjugates $G_a$ annihilation operators of a pair of oscillator algebras. There is no finite-dimensional matrix representation for this algebra which can be seen by taking the matrix trace of both sides. We can also define the number operator
\bea
N &=& G_a G^a
\eea
which counts the number of creation operators $G^a$ minus the number of annihilation operators $G_a$,
\ben
[N,G^a] &=& G^a\cr
[N,G_a] &=& -G_a\label{osc2}
\een
We assume that there is a ground state annihilated by all $G_a$,
\bea
G_a\left|0,0\right> = 0
\eea
and we define orthonormal states as
\bea
\left|m,n\right> &=& \frac{1}{\sqrt{m!n!}}(G^1)^m (G^2)^n \left|0,0\right>
\eea
Ignoring the normalization, we may also write such states in the form
\bea
G^{a_1}\cdots G^{a_{N-1}} \left|0,0\right>
\eea
This shows that these states are symmetric under permutations of the $a_i$'s. They correspond to a Young tableaux with one row with $N-1$ boxes and constitute the $N$-dimensional representation of $SU(2)$ \cite{Mathur}. Thus we have states $\left|m,N-m-1\right>$ ($m=0,...,N-1$) in dimension-$N$ representation of $SU(2)$. Since these states are characterized by one integer $m$ (given the dimension $N$), we will henceforth abbreviate the notation, and write these states as $\left|m\right>$, or as $\left|m,N \right>$ if we want to indicate the dimension $N$.

The Schwinger construction of $SU(2)$ algebra means that we define
\bea
K_I &=& G^a (\sigma_I)_a{}^b G_b
\eea
as an operator\footnote{One may think that $G_b (\sigma_I)_a{}^b G^a$ is another operator, but it is not since by commuting the $G's$ we just produce $\delta^a_b (\sigma_I)_a{}^b = 0$.} and we have a representation as 
\bea
\left<m,N\right| K_I \left|n,N\right> &=& (K_I)_{m}{}^{n}
\eea
If we take $N=2$ this will reproduce the Pauli matrices, and in general we get dimension-$N$ representation of $SU(2)$. Since the $G^a$ increase $N$ by one unit, we obtain the TGRVV matrices that connects $SU(2)$ representations $N-1$ and $N' = N$ respectively, as
\bea
\left<m',N\right| G^a \left|m,N-1\right> &=& (G^a)_{m'}{}^{m}
\eea
For example, taking $N'=3$ we get states $\left|1,0\right>,\left|0,1\right>$ and with $N=2$ we get states $\left<2,0\right|, \left<1,1\right|,\left<0,2\right|$ and then we get
\bea
\left<2,0\right|G^1\left|1,0\right> &=& \sqrt{2}\cr
\left<1,1\right|G^1\left|0,1\right> &=& 1
\eea
and
\bea
\left<0,2\right|G^2\left|0,1\right> &=& \sqrt{2}\cr
\left<1,1\right|G^2\left|1,0\right> &=& 1
\eea
and all other entries vanish. We have obtained the matrix representation
\bea
G^1 = \(\begin{array}{cc}
\sqrt{2} & 0\\
0 & 1\\
0 & 0
\end{array}\), \qquad G^2 = \(\begin{array}{cc}
0 & 0\\
1 & 0\\
0 & \sqrt{2}
\end{array}\)
\eea
The extension to arbitrary $N-1$ and $N'=N$ is provided by the TGRVV matrices
\bea
(G^1)_{m'}{}^{n} = \(\begin{array}{ccc}
\sqrt{N-1} & \cdots & 0\\
  & \ddots & \\
0 & \cdots & \sqrt{1}\\
0 & \cdots & 0
\end{array}\), \qquad (G^2)_{m'}{}^n = \(\begin{array}{ccc}
0 & \cdots & 0\\
\sqrt{1} & \cdots & 0\\
     & \ddots & \\
0  & \cdots & \sqrt{N-1}
\end{array}\) 
\eea
These matrices satisfy the TGRVV three-algebra
\bea
[G^a,G^b;G^c] &=&  f^{ab}{}_{cd}G^d
\eea
where the structure constants are given by 
\bea
f^{ab}{}_{cd} &=& -2\delta^{ab}_{cd}
\eea
and where the three-bracket is defined as
\bea
[G^a,G^b;G^c] &=& G^a G_c G^b - G^b G_c G^a
\eea
This algebra can be showed directly from the oscillator algebra without assuming a matrix representation. Matrix realization is obtained by inserting a complete set of states between the operators and by using $\left<m',N'\right|G^a\left|m,N\right> = 0$ unless $N'=N+1$. Thus we may expand out the three-bracket in terms of commutators as
\bea
[G^a,G^b;G^c] &=& G_c[G^a,G^b] + [G^a,G_c] G^b - [G^b,G_c] G^a
\eea 
and then apply the oscillator algbra, and we immediately arrive at the TGRVV three-algebra.

We may extend the TGRVV algebra to a larger three-algebra by including elements of the form
\ben
G^{a_1} G_{b_1} G^{a_2} \cdots G_{b_{k-1}} G^{a_k}\label{three-algebra}
\een
We note that this operator increases $N$ by one unit, just like $G^{a_1}$ does in the case when $k=1$. Using matrix realization of the $G's$ we see that this alternating multiplication structure is the most natural one and means that we multiply $N\times (N-1)$ matrix with $(N-1) \times N$ matrix alternatingly and it gives us again an $N \times (N-1)$ matrix. However not all of these elements are independent. To find the independent generators, we have to extract the number operators. If we multiply a three-algebra generator by a number operator from the right, it will obey the same algebra as the generator without that additional number operator,
\bea
[T^a N,T^b;T^c] &=& [T^a,T^b;T^c] N\cr
[T^a,T^b N;T^c] &=& [T^a,T^b;T^c] N\cr
[T^a,T^b;T^cN] &=& [T^a,T^b;T^c] N
\eea
This is not entirely obvious since the number operator actually counts number of $G^a$ minus number of $G_a$ standing to the right. But then we note that the combination $T_c T^b$ always contains an equal number of $G^a$ and $G_a$. So in effect $N$ can always be pulled out of any three-bracket. Moreover, the ordering of the labels $a_1,...,a_k$ does not matter since we can use the oscillator algebra to relate any two orderings by adding generators of lower ranks. The same is true for the ordering of $b_1,...,b_{k-1}$. The number of such symmetric tensors is $(k+1) k$. Removing traces means we remove $(k-1)k$ components thus leaving us with $2k$ independent components. If we then let $k$ to run over $k = 1,...,N-1$ we find in total
\bea
\sum_{k=1}^{N-1} 2k &=& (N-1)N
\eea
independent three-algebra generators. This coincides with the number of elements in a generic $N\times (N-1)$ matrix. 

The map from products of $G^a$ and $G_b$'s into a bifundamental matrix (three-algebra generator) is provided by 
\bea
\left<n',N\right| G^{a_1} G_{b_1} G^{a_2} \cdots G_{b_{k-1}} G^{a_k} \left|n,N-1\right> &=& (T^a)_{n'}{}^n
\eea
All the generators (\ref{three-algebra}) are linearly independent after we have extracted the number operators. It implies that any $N\times (N-1)$ bi-fundamental matrix $M_{n'}{}^n$ can be obtained as some linear combination of these three-algebra generators,
\bea
M_{n'}{}^n &=& \sum_{a=1}^{N(N-1)} c_a (T^a)_{n'}{}^n
\eea
Here $a$ is a collective index which is associated with the sequence $a_1,b_1,...,b_{k-1},a_k$ modulo number operators.

From the oscillator algebra (\ref{osc1}) we can compute the following matrix elements
\ben
\<m',N\! G^a G_a\|n',N\> &=& (N-1)\delta_n^{n'}\cr
\<m,N-1\! G_a G^a\|n,N-1\> &=& N\delta_m^n\label{radii}
\een
where we used 
\bea
G_a G^a &=& 2 + G^a G_a
\eea
We may also note that the above is consistent with cyclicity of trace,
\bea
\tr_N(G^a G_a) = (N-1)N = \tr_{N-1}(G_a G^a)
\eea

We have the following commutation relations
\bea
[J_I,G^a] &=& G^b(\sigma_I)_b{}^a\cr
[J_I,G_a] &=& -(\sigma_I)_a{}^b G_b
\eea
from which it follows that we have the following operator identity \cite{Nastase:2009ny}
\ben
G^a J_{(I_1}\cdots J_{I_k)} G_a &=& \(G^a G_a - k\) J_{(I_1}\cdots J_{I_k)}\label{contract}
\een
These results will be very useful when we consider the Higgs mechanism \cite{Mukhi:2008ux} by following \cite{Nastase:2009ny}.

\subsection{The associated gauge group}
So far we have obtained the three-algebra generators in Eq (\ref{three-algebra}). It remains to obtain the associated Lie algebra or the gauge group. The smallest three-algebra is generated by $G^a$ and has associated Lie algebra generators
\bea
J &=& [\cdot,G^a;G^a]\cr
J_I &=&  [\cdot,G^a;G^b] (\sigma_I)_a{}^b
\eea
By using the Fierz identity (which we derive in Appendix $A$)
\bea
(\sigma_I)_c{}^d (\sigma_J)_a{}^b - (\sigma_J)_c{}^d (\sigma_I)_a{}^b &=& i \epsilon_{IJK} \((\sigma_K)_c{}^b \delta_a^d - (\sigma_K)_a{}^d \delta_c^b\)
\eea
the hermitian fundamental identity and the TGRVV algebra, we can obtain the commutation relations
\bea
[J_I,J_J] &=& 2i \epsilon_{IJK} J_K\cr
[J_I,J] &=& 0
\eea
Here the multiplication of generators is by composition of maps, thus $J_I J_J (X) := J^I (J^J(X))$.

We can also compute 
\bea
J^I J^I (G^a) &=& 3 G^a
\eea
which is in accordance with that $G^a\|0,0\>$ constitute the $N=2$ fundamental representation of $SU(2)$. Higher-dimensional representations are obtained by acting on $G^{a_1}...G^{a_k}\|0,0\>$. For instance $G^{a_1} G^{a_2}\|0,0\>$ gives $N=3$ adjoint representation of dimension $N^2-1 = 8$ by the following computation
\bea
J^I J^I (G^a G^b) &=& J^I J^I (G^a) G^b + 2 J^I(G^2) J^I(G^b) + G^a J^I J^I(G^b)
\eea
where we notice that in general
\bea
[XY,G^a;G^b] &=& [X,G^a;G^b]Y + X[Y,G^a;G^b] - X [G_b,G^a] Y
\eea
and then we make use of the oscillator representation of the TGRVV algebra together with $(\sigma^I)_a{}^b \delta^a_b = 0$. Now this could lead us to conclude that the gauge group associated to the smallest three-algebra must be $SU(2) \times U(1)$. But as we will now show, this is not the only possible choice of gauge group.

\subsection{Gauge group $U(2) \times U(1)$}
In the minimal case we have two three-algebra generators 
\bea
T^a &=& \sqrt{\frac{2\pi}{K}} G^a
\eea
for $a=1,2$. These generate the three-algebra 
\bea
[T^a,T^b;T^c] &=& f^{ab}{}_{cd} T^d
\eea
with structure constants
\bea
f^{ab}{}_{cd} &=& \frac{2\pi}{K} \(\delta^a_d \delta^b_c - \delta^a_c \delta^b_d\)
\eea
We will now proceed to find the associated gauge group. Let $(\sigma_A)^b{}_a$ denote the Pauli sigma matrices, which satisfy
\bea
\sigma_A\sigma_B &=& \delta_{AB} + 2i \epsilon_{ABC} \sigma_C
\eea
We then define 
\bea
G_{AB} &=& (\sigma_A)^b{}_a (\sigma_B)^a{}_b
\eea
and $G^{AB}$ as its inverse. Here we get
\bea
G_{AB} &=& 2\delta_{AB}\cr
G^{AB} &=& \frac{1}{2} \delta^{AB}
\eea
and we find
\bea
G^{AB} (\sigma_A)^b{}_a (\sigma_B)^c{}_d &=& \delta^b_d \delta^c_a - \frac{1}{2} \delta^b_a \delta^c_d
\eea
as follows from a Fierz identity that we derive in the Appendix $A$. Structure constants of $SU(2) \times U(1)^r$ are given by
\bea
f^{bc}{}_{da} &=& -\frac{2\pi}{K} G^{AB} (\sigma_A)^b{}_a (\sigma_B)^c{}_d + \lambda \delta^b_a \delta^c_d
\eea
which for antisymmetry in $bc$ requires 
\bea
\lambda &=& \frac{\pi}{K}
\eea
This in turn implies that the $U(1)^r$ Chern-Simons levels $k_i$ are constrained by
\bea
2\pi \sum_{i=1}^r \frac{1}{k_i} &=& -\frac{\pi}{K}
\eea
If $r=2$ we have a solution
\bea
k_1 &=& 2K\cr
k_2 &=& -K
\eea
and if $r=1$ we have the solution
\bea
k_1 &=& -2K
\eea
These solutions correspond to gauge groups $U(2)_K \times U(1)_{-K}$ and $SU(2)_K \times U(1)_{-2K}$ respectively. We note that 
\bea
U(2)_K &=& SU(2)_K \times \(U(1)/{\mb{Z}_2}\)_{2K}
\eea
In general $g \in SU(N)$ means det($g$) $=1$ and $g \rightarrow e^{\frac{2\pi i}{N}} g$ is in fact an $SU(N)$ rotation. This means that $U(1)$ inside $U(N)$ shall only act by $e^{i\alpha}$ where $\alpha \sim \alpha + \frac{2\pi}{N}$, thus $U(1)/{\mb{Z}_N}$.

For $U(2)_K\times U(1)_{-K}$ the Chern-Simons terms are given by
\bea
&& \frac{K}{4\pi} \tr_{SU(2)} \(A dA + \frac{2i}{3} A^3\) + \frac{K}{4\pi} \(2A^1 dA^1 - A^2 dA^2\)\cr
&=& \frac{K}{4\pi} \tr_{U(2)} \(A dA + \frac{2i}{3} A^3\) - \frac{K}{4\pi} A^2 dA^2
\eea
where in the second line $A^b{}_a = A^I (\sigma_I)^b{}_a + A^1 \delta^b_a$.
and the covariant derivative is 
\bea
D_{\mu} Z_a &=& \partial_{\mu} Z_a - i A_{\mu}^I (\sigma_I)^b{}_a Z_b - i \(A^1_{\mu} + A^2_{\mu}\) Z_a\cr
&=& \partial_{\mu} Z_a - i A_{\mu}{}^b{}_a Z_b - i A^2_{\mu} Z_a
\eea

We expect this will generalize to higher values on $N$. Starting with the abstractly defined three-algebra operators (\ref{three-algebra}) where $k = 0,...,N-1$, we expect to find three-algebra structure constants generalizing those of the TGRVV algebra, and which uniquely corresponds to the gauge group $U(N)_K \times U(N-1)_{-K}$. We note that the other case of $SU(2) \times U(1)$ gauge group is very special to the case when $N=2$. This Lie algebra does not generalize to arbitrary $N$ as was shown in \cite{Gustavsson:2010yr}.

\section{The novel Higgs mechanism}\label{4}
In ABJM theory with gauge group $U(N) \times U(N-1)$ there are two gauge fields. To be specific, let us consider the gauge field $A^L$ which is associated with $U(N)$. We can expand this gauge field in fuzzy spherical harmonics as
\bea
A^L &=& \sum_{\l = 0}^{N-1} \sum_{m=-\l}^{\l} A^L_{\l m} Y_{\l m}(J_I)
\eea
where 
\bea
Y_{\l m}(J_I) &=& J_{(I_1} \cdots J_{I_\l)} - traces
\eea
Here the $J_I$ denote the $SU(2)$ Lie algebra generators in the $N$-dimensional representation. This is then the gauge field of $U(N)$ gauge group. Deconstruction means that we interpret this $U(N)$ gauge field as a $U(1)$ gauge field on a fuzzy $S^2$ with radius $r$,
\bea
A^L &=& \frac{(N^2-1)^{\frac{\l}{2}}}{r^{\frac{\l}{2}}} A^L_{\l m} Y_{\l m}(x_I)
\eea
where the spherical harmonics are star-multiplied. 

For the scalar field, when it comes to the Higgs mechanism and zero magnetic flux sector, all we need to do, is just to insert its vacuum expectation value
\ben
Z^A &=&  \sqrt{m\hbar} \(\begin{array}{c}
G^a\\
0
\end{array}\)\label{vev}
\een
Fluctuations around this vacuum will also be important on their own, but they should not be considered in the Higgs mechanism itself. 

As far as the novel Higgs mechanism concerns, all we need to do is to evaluate the scalar field kinetic term $\L_{kin}$ on this vacuum expectation value, and then add the Chern-Simons term $\L_{CS}$. We then solve for the gauge field $B_{\mu}$ from the classical gauge field equation of motion and plug back into the Lagrangian. In the process, the vacuum expectation value of the scalar fields will not make the gauge field massive as usual in the Higgs mechanism. Instead it will make a non-dynamical gauge field $A_{\mu}$ dynamical. This is why this Higgs mechanism is refered to as novel. 

\subsection{The kinetic term}
For the kinetic term we will closely follow the computation in \cite{Nastase:2009ny}. We evaluate $\L_{kin}$ on the vacuum expectation value (\ref{vev}). We then get 
\bea
\L'_{kin} &=& -m \tr_N \(A^L_{\mu} G^a G_a A^{L,\mu} + G^a A^R_{\mu} A^{R,\mu} G_a + 2 A^L_{\mu} G^a A^{R,\mu} G_a\)
\eea
Here
\bea
\tr_N(\cdots) &=& \sum_{m'=0}^{N-1} \<m',N\! \cdots \|m',N\>
\eea
What is important to note here, is that the rising operators $G^a$ connect states of different dimensionality (being a rising operator, it rises the dimensionality of a state by one unit). When we use cyclicity of trace, $\tr_N$ may thus turn into $\tr_{N-1}$ and vice versa. We have for example 
\bea
\tr_N \(G^a A^R A^R G_a\) &=& \tr_{N-1} \(A^R G_a G^a A^R\)
\eea
Using trace properties of the TGRVV matrices as derived in the appendix, we get
\bea
\L'_{kin} &=& -m \((N-1) \tr_N (A^L A^L) + N \tr_{N-1} (A^R A^R) + 2\tr_N \(A^L G^a A^R G_a\)\)
\eea
To get further we use that
\bea
G^a J_{(I_1}\cdots J_{I_{\l})} G_a &=& (G^a G_a - \l) J_{(I_1}\cdots J_{I_{\l})} 
\eea
as an operator identity. Then the last term becomes
\bea
2 \sum_m \<m',N\! A^L (G^a G_a - \l) J_{(I_1}\cdots J_{I_{\l})} \|m',N\> A^R_{I_1\cdots I_{\l}} &=& 2(N-1-\l)\tr_N \(A^L A^R_{\l}\) 
\eea
where $A^R_{\l}$ is the gauge field up-lifted from $U(N-1)$ to $U(N)$ simply by replacing $J_I$ in $(N-1)$-dimensional representation of $SU(2)$ with $J_I$ in $N$-dimensional representation. We now map $J_I$ into $x_I$, and get
\bea
\L'_{kin} &=& -m \int_{S^2} \((N-1)N \(A^L A^L + A^R A^R\) + 2N (N-1-\l) A^L A^R_{\l} \)
\eea
We relate $A^R_{\l}$ in dimension $N$ to $A^R$ in dimension $N-1$ simply by rescaling 
\bea
A^R_{\l} &=& \(\frac{N^2-1}{(N-1)^2-1}\)^{\frac{\l}{2}} A^R
\eea
The star-product shall also be corrected in a similar when changing $N$ to $N-1$, but the star-product is not needed for the inner products and so it will not be of any concern to us here. We then use the Taylor expansion 
\bea
\(\frac{N^2-1}{(N-1)^2-1}\)^{\frac{\l}{2}} (N-1-\l) &=& N-1-\frac{\l (\l+1)}{2N} + \O\(\frac{1}{N^2}\)
\eea
and find the last term as
\bea
-m 2N(N-1) \int_{S^2} \(A^L A^R - \frac{1}{2N(N-1)} A^L \hat \Box A^R\)
\eea
and our final result for the kinetic term is
\bea
\L'_{kin} &=& -\frac{m}{2} \int_{S^2} \( N(N-1)\(A^L + A^R\)^2 - A^L \hat \Box A^R\)
\eea
Here we have noted that 
\bea
\hat{\Box} Y_{lm} &=& l(l+1) Y_{lm}
\eea
and we define
\bea
\hat \Box &=& r^2 \Box,\cr
\Box &=& G^{mn} D_m \partial_n
\eea
Here
\bea
r &=& \frac{R}{2}
\eea
denotes the radius of the $S^2$ base manifold \cite{Nastase:2009ny}.

\subsection{The Chern-Simons term}
The Chern-Simons term is
\bea
\L'_{CS} &=& \frac{1}{2}\(\tr_{N} A^L dA^L - \tr_{N-1} A^R dA^R\)
\eea
We map $\tr_N$ onto $N\int_{S^2}$ as before, and we get
\bea
\L'_{CS} &=& \frac{N}{2} \int_{S^2} \(A^L dA^L - A^R dA^R\) + \frac{1}{2} \int_{S^2} A^R dA^R
\eea

\subsection{Integrating out $B$}
If we define
\bea
B &=& A^L+A^R\cr
A &=& A^L-A^R
\eea
then the sum $\L_{CS} + \L_{kin}$ on the scalar field vacuum expectation value, becomes
\bea
\frac{K}{2\pi} \int_{S^2} \[\frac{N}{4}\epsilon^{\mu\nu\lambda} B_{\mu} F_{\nu\lambda} + \frac{1}{8} \epsilon^{\mu\nu\lambda} A_{\mu}\partial_{\nu}A_{\lambda} - m N(N-1) B_{\mu}B^{\mu} + \frac{m}{4} \(B_{\mu}\hat\Box_{S^2} B^{\mu} - A_{\mu} \hat\Box_{S^2} A^{\mu}\)\]
\eea
where we have used
\bea
A^R dA^R &=& \frac{1}{4}\(BdB - 2AdB + AdA\)
\eea
and where we have suppressed the terms $BdB$, $AdB$ and $B\Box B$ which will be of order $\O(\frac{1}{N})$. We can now solve algebraically for $B$,
\bea
B^{\mu} &=& \frac{1}{8m(N-1)} \epsilon^{\mu\nu\lambda} F_{\nu\lambda}
\eea
In order to get the correct normalization of the graviphoton term we now have to change the normalization of the gauge field as 
\bea
A_{\mu} &=& 2 A'_{\mu}
\eea
and then we drop the prime not to clutter the final result. We then get
\bea
-\frac{K}{4\pi} \int_{S^2} \epsilon^{\mu\nu\lambda} A_{\mu}\partial_{\nu} A_{\lambda} + \frac{K}{16\pi^2 R} \int_{S^2} d^2 \sigma \sqrt{G} \(F_{\mu\nu}^2 - \frac{2}{R^2} A_{\mu} \Box_{S^2} A^{\mu}\) + \O\(\frac{1}{N}\)
\eea
We make an integration by parts
\bea
-A_{\mu} \Box_{S^2} A^{\mu} &=& \partial_m A_{\mu} \partial^m A^{\mu}
\eea
This term combines with other terms into $F_{m\mu} F^{m\mu}$, which in turn combines with other terms into a five-dimensional Maxwell term $F_{MN} F^{MN}$ where $M = (\mu,m)$. The first term can be rewritten as
\bea
\frac{1}{8\pi^2} \int A \wedge F \wedge W
\eea
where
\bea
W &=& \frac{K}{2r^2} \Omega_S
\eea
and $\Omega_S$ denotes the volume form on $S^2$ of radius $r=\frac{R}{2}$. We can write 
\bea
W &=& dV
\eea
and $V$ will be our graviphoton field. Locally $V$ is the gauge potential of a magnetic monopole of strength $K$.

From the Maxwell term we read of the gauge coupling constant $g_{YM}^2 = 4\pi^2 (R/K)$. This corresponds to the dimensional reduction along the Hopf fiber of radius $R/K$ of the M5 brane coupling constant. 

The generalization to non-abelian sYM is straightforward. We then instead start with ABJM gauge group $U(NM) \times U((N-1)M)$ and three-algebra generators are taken as $T^a \otimes T^{a'}$ where $T^a$ are as before constructed out of alternating products of TGRVV operators, and $T^{a'}$ are three-algebra generators associated with $U(M)\times U(M)$ Lie algebra. Then the tensor product of such three-algebra generators will generate $U(NM) \times U((N-1)M)$ Lie algebra. In particular, if we realize the generators by matrices, the tensor product generators $(T^{aa'})_{ik}^{jl} := (T^a)_i^j (T^{a'})_k^l$ will be $NM\times (N-1)M$ matrices. In the deconstructed theory we map $\tr_{NM}$ into $N \int_{S^2} \tr_M$ and we descend to $U(M)$ sYM.

\section{Partition functions for zero modes}\label{5}
The graviphoton term enable us to show that the zero mode contribution to the partition functions of D4 brane on $\mb{R}^{1,2}\times S^2$ matches with the zero mode contribution to the M5 brane partition on $\mb{R}^{1,2} \times (S^3/\mb{Z}_K)$. By zero modes we mean field configurations that extremize the classical action, thus are solutions to the classical field equations of motion. There will of course also be quantum fluctuations around these classical solutions, but we will not consider their contributions here. We will now proceed to compute the zero mode contributions to the partition functions of D4 and M5 branes.

\subsection{D4 brane}
If we complete the deconstruction of the D4 brane theory, and we may assume this is on $\mb{R} \times T^2 \times S^2$, we will in particular find the following piece in the Lagrangian 
\bea
\sqrt{G}\L &=& -\frac{1}{2} \left| F- \frac{2}{R} \Phi \Omega_S \right|^2 - \frac{R}{2} V \wedge F \wedge F - \frac{1}{2}\left| d\Phi\right|^2
\eea
where $\Phi$ denotes one of the five scalar field. (As it will turn out, the overall normalization of this Lagrangian will play no role in our result for the zero modes, and so we will be ignorant about this factor.) Zero modes can not arise in any of the other terms in the Lagrangian and so we find no need to write out those terms here. (The full Lagrangian is found in \cite{Gustavsson:2011af}). Here $\Omega_S$ denotes the volume form on $S^2$ and $\Omega_M$ is defined by either on of the relations
\bea
*\Omega_S &=& \Omega_M\cr
*1 &=& \Omega_M \wedge \Omega_S
\eea
where the Hodge-star is defined with respect to five-dimensional space-time. If we define 
\bea
G &=& F - \frac{2}{R}\Phi \Omega_S
\eea
as a modified field strength subject to the Bianchi identity
\bea
dG &=& -\frac{2}{R} d\Phi \wedge \Omega_S
\eea 
then the Lagrangian gives the equations of motion
\ben
d^{\dag} G &=& *(G \wedge \Omega_S)\label{ones}\\
\triangle \Phi &=& *(G \wedge \Omega_M)\label{twos}
\een
In \cite{Lambert:2011eg}, \cite{Lambert} it was claimed that the graviphoton term constrains $G \wedge \Omega_S = 0$. To show this, let us decompose
\bea
G &=& A + B + C
\eea
where 
\bea
A &=& f\Omega_S
\eea
and 
\bea
B \wedge \Omega_M &=& 0\cr
C \wedge \Omega_S &=& 0\cr
C \wedge \Omega_M &=& 0
\eea
and more specifically
\bea
B &=& B_{\mu\nu} dx^{\mu} \wedge dx^{\nu}\cr
C &=& C_{\mu m} dx^{\mu} \wedge d\sigma^m
\eea
where $\mu$ are vector indices on $\mb{R} \times T^2$ and $m$ are vector indices on $S^2$. 
Then (\ref{ones}) becomes 
\bea
d\(f \Omega_M + *(B+C)\) &=& B \wedge \Omega_S\cr
\eea
Since left-hand side is exact, as well as $\Omega_S$ is closed, we have locally
\bea
B &=& dE
\eea
where $E = E_{\mu}(x^{\nu}) dx^{\mu}$ is a one-form. Then we get
\bea
f \Omega_M + *(B+C) &=& E \wedge \Omega_S + d(...)
\eea
Index structures on all these terms are different and therefore $E \wedge \Omega_S = d(...)$, and that means $E = de(x^{\mu})$ but then $B = dE = dde = 0$ and so $G\wedge \Omega_S = 0$ follows as it was claimed in \cite{Lambert:2011eg}, \cite{Lambert}. Thus the zero modes are the harmonic two-forms, 
\bea
F &=& f \Omega_S 
\eea
where $f$ is constant. The equations of motion then reduce to
\bea
\triangle \(f - \frac{2}{R} \Phi\) &=& 0\cr
-\frac{2}{R} \(f - \frac{2}{R} \Phi\) + \triangle \Phi &=& 0
\eea
Solutions to these equations are given by
\bea
F &=& \frac{2 n}{R^2} \Omega_S \cr
\Phi &=& \frac{n}{R} 
\eea
where $n \in \mb{Z}$. We note that
\bea
\int \Omega_S &=& 4\pi \(\frac{R}{2}\)^2\cr
&=& \pi R^2
\eea
so that 
\bea
\int F &=& 2\pi n
\eea
Plugging these solutions back into the Lagrangian gives
\bea
\sqrt{G} \L &=& 0
\eea

\subsection{M5 brane}
Here we will use the action which was deconstructed in \cite{Gustavsson:2011af} of M5 brane on $\mb{R} \times T^2 \times S^3$. We now also define a field strength three-form as
\bea
H &=& \frac{1}{6} H_{\alpha\beta\gamma} d\sigma^{\alpha} \wedge d\sigma^{\beta} \wedge d\sigma^{\gamma} + \frac{1}{2} H_{\mu\alpha\beta} dx^{\mu} \wedge d\sigma^{\alpha} \wedge d\sigma^{\beta}
\eea
where indices $\mu = 0,1,2$ are associated with $\mb{R} \times T^2$, and $\alpha=\theta,\varphi,\psi$ are associated with spherical coordinates $\sigma^{\alpha}$ on $S^3$. From this we note that $H_{\mu\nu\lambda}$ and $H_{\mu\nu\alpha}$ are absent. This is a Lorentz non-covariant formulation of the M5 brane. It means that we have no flux on $T^2$. Even if we would introduce such component, we would still have one components on $S^3$. Since there are no harmonic one-forms on $S^3$, this would not be a harmonic zero mode. We can immediately conclude that we would have a trouble to match with the zero mode contribution of the D4 brane if we had a non-vanishing magnetic flux $\int G \wedge \Omega_S = \int_{T^2} F$ through $T^2$. It is therefore fortunate for us that the graviphoton term constrains this flux to vanish in the D4 brane. 

Let us now proceed with the detailed computation. Let us recycle the notation and here denote by $\Omega_S$ the volume form on $S^3$, and define $\Omega_M$ through the relation
\bea
* \Omega_S &=& -\Omega_M
\eea
Let us also define the three-form gauge field strength as
Then the Maxwell type Lagrangian that was found in \cite{Gustavsson:2011af} can be written in the form 
\bea
\sqrt{g} \L &=& -\frac{1}{2} \left| H + \frac{2}{R} Y \Omega_S \right|^2 - \frac{1}{2} \left| dY \right|^2
\eea
(Here $g$ denotes the determinant of the metric on $\mb{R}^{1,2}\times S^3$.) From this Lagrangian we derive the equations of motion
\bea
d^{\dag} \(H + \frac{2}{R}Y \Omega_S\) &=& 0
\eea
If we now assume that the field strength is a zero mode, then it must be on the form
\bea
H &=& h \Omega_S
\eea
since there is no harmonic two-form on $S^3$. We then get the scalar field equation of motion
\bea
\triangle Y + \frac{2}{R} \(h + \frac{2}{R}Y\) &=& 0
\eea
Solutions to these equations of motion are given by
\bea
H &=& \frac{n}{\pi R^3} \Omega_S\cr
Y &=& -\frac{n}{2\pi R^2}
\eea
We note that
\bea
\int_{S^3} \Omega_S &=& 2\pi^2 R^3
\eea
and so we have
\bea
\int H &=& 2\pi n
\eea
Inserting these solutions into the Lagrangian gives
\bea
\sqrt{g} \L &=& 0
\eea

\subsection{Matching the partition functions}
The zero mode contribution the partition function is given by
\bea
\Z_{zero-modes} &=& \sum_n e^{i S_{classical}}
\eea
where 
\bea
S_{classical} &=& \int \L
\eea
Since we found $\L = 0$ on all the zero mode solutions, we see that the zero modes contribute
\bea
\Z_{zero-modes} &=& \sum_n 1
\eea
both to the M5 and the D4 brane partition functions. 

We may also assign a physical interpretation to the integer number $n$ that appears in both M5 and D4. This integer number counts the number of M2 branes \cite{Gustavsson:2011af}, \cite{Lambert:2011eg}. So we shall keep this number $n$ fixed and equal for M5 and D4. In this case the zero mode contribution reduces to a finite sum, consisting of a single number $n$ and we get
\bea
\Z_{zero-modes} &=& 1
\eea
for both M5 and D4.

\section{Discussion} 
We have seen that $G^a$ can be thought of as creation operators. This opens up a fascinating possibility of considering an enlarged three-algebra containing product such as $G^a G^b$ and so on, which enforces us to consider matrices of type $N \times (N-2)$ and so on. We may need all kinds of rectangular matrices $N \times M$ where $M=1,2,...,(N-1)$. If we consider a space of states which is sum $\|m_1,1\> + \|m_2,2\> + \cdots + \|m_N,N\>$, it is clear that we will form a closed finite-dimensional three-algebra. The smallest non-trivial such three-algebra is generated by $G^a$, $G^a G^b$ acting on states $\|m_1,1\> + \|m_2,2\> + \|m_3,3\>$. Non-vanishing matrix elements are for instance
\bea
\<m',3\!G^a G^b \|m,1\>
\eea
and since there is no state $\|m,4\>$ we do not get any non-vanishing matrix elements for $G^a G^b G^c$ and therefore the algebra will be finite-dimensional. The three-algebra generated by the operators is infinite-dimensional, but when we wedge it with these states, it become the finite-dimensional three-algebra of matrices $G^a$ of types $2\times 1$ and $3\times 2$ and matrices $G^a G^b$ of type $3\times 1$. We can then compute a three-bracket 
\bea
[G^a,G^b;G^c G^d] &=& G^a G_d G_c G^b - G^b G_d G_c G^a
\eea
where $G_d G_c$ is a $1\times 3$ matrix, which can multiply $G^a$ of size $2\times 1$ from the left, and on the right we can multiply be $G^b$ of size $3\times 2$ for the first term. For the second term we instead take $G^b$ of size $2\times 1$ and $G^a$ of size $3\times 2$. This sounds strange but if we think on $G^a$ as operators acting on states, then what we do is nothing but putting the three-bracket inside two states, and evaluating the resulting matrix elements. Thus for the first term we compute
\bea
&& \(\<m'_1,1\! + \<m'_2,2\! + \<m_3,3\!\) G^a G_d G_c G^b \(\|m_1,1\> + \|m_2,2\> + \|m_3,3\>\) \cr
&=& \<m'_2,2\! G^a G_d G_c G^b \|m_3,3\>\cr
&=& \sum_{n,p}\<m'_2,2\! G^a \|n,1\>\<n,1\! G_d G_c \|p,3\>\<p,3\! G^b \|m_3,3\>
\eea
It would be interesting to find the associated gauge group of this three-algebra and check if the $N^{3/2}$ scaling is manifest. In any case, this is a very natural way of quantizing $S^3$, by including operators $G^a G^b$ and all those, thus extending those of alternating form in (\ref{three-algebra}). Namely, on the function side we want to generate all the spherical harmonics on $S^3$. We do this by precisely this extension. In an embedding of $S^3$ in ${\mb{R}}^4$ with euclidean coordinates $x^i$ constrained by $x^i x^i = R^2$, the spherical harmonics are the symmetric traceless functions $x^{i_1} \cdots x^{i_k}$. Then we just switch to the complex basis $G^a$ and we see that generators (\ref{three-algebra}) just are not enough, as these all come with the same phase factor $e^{i\psi}$ along the Hopf fiber whose coordinate we denote $\psi$. To get all phases $e^{in\psi}$ we need to consider more general class of operators than those alternating series. We need to include $G^a G^b$ and such operators. Clearly the framework with matrices of fixed range is too limited for this purpose. Our suggestion is that we may consider a sum of matrices of all times $N\times M$ for $M=1,...,N-1$.


\subsubsection*{Acknowledgements}
This work was supported by NRF Mid-career Researcher Program 2011-0013228.

\newpage

\appendix

\section{The Pauli sigma matrices}
The Pauli sigma matrices obey the algebra
\bea
\sigma_I \sigma_J &=& \delta_{IJ} + i \epsilon_{IJK} \sigma_K
\eea
For anticommuting spinors we have the Fierz identity
\bea
\chi \psi^{\dag} &=& -\frac{1}{2} \(\psi^{\dag}\chi + (\psi^{\dag}\sigma_I \chi)\) \sigma_I
\eea
We will use the index convention
\bea
\tr(\chi \psi^{\dag}) &=& \chi_a \psi^a
\eea
Thus, writing out indices, we have
\bea
\chi_a \psi^b &=& -\frac{1}{2} \(\psi^c \chi_c \delta_a^b + (\psi^c (\sigma_I)_c{}^d \chi_d) (\sigma_I)_a{}^b\)
\eea
We can verify this identity by contracting indices by $\delta^a_b$ and by $(\sigma_J)_b{}^a$ respectively.

We use the Fierz identity to get
\bea
\psi_1^{\dag} \sigma_I \chi_1 \psi_2^{\dag} \sigma_J \chi_2 &=& -\frac{1}{2} \psi_1^{\dag} \sigma_I \sigma_J \chi_2 \psi_2^{\dag} \chi_1 - \frac{1}{2} \psi_1^{\dag} \sigma_I \sigma_K \sigma_J \chi_2 \psi_2^{\dag} \sigma_K \chi_1
\eea
We next obtain the identity
\bea
\sigma_I \sigma_K \sigma_J &=& \delta_{IK} \sigma_J + \delta_{JK} \sigma_I - \delta_{IJ} \sigma_K - i \epsilon_{IJK}
\eea
and we get
\bea
2(\sigma_I)_a{}^d (\sigma_J)_c{}^b &=& \delta_{IJ} \delta_a^b \delta_c^d + i \epsilon_{IJK} \((\sigma_K)_a{}^b \delta_c^d - \delta_a^b (\sigma_K)_c{}^d\)\cr
&&+ (\sigma_J)_a{}^b (\sigma_I)_c{}^d + (\sigma_I)_a{}^b (\sigma_J)_c{}^d - \delta_{IJ} (\sigma_K)_a{}^b (\sigma_K)_c{}^d
\eea
where the overall minus sign is from anticommuting fermions. We contract $IJ$ and get
\bea
2(\sigma_I)_a{}^d (\sigma_I)_c{}^b &=& 3 \delta_a^b \delta_c^d - (\sigma_I)_a{}^b (\sigma_I)_c{}^d
\eea
Then by making the most general ansatz compatible with $SU(2)$ covariance,
\bea
(\sigma_I)_a{}^b (\sigma_I)_c{}^d &=& a \delta_a^d \delta_c^b + b \delta_a^d \delta_c^d + c \epsilon_{ac}\epsilon^{bd}
\eea
(and we may note that the last term is not independent of the first two) we can determine the coefficients as 
\bea
(\sigma_I)_a{}^b (\sigma_I)_c{}^d &=& 2\delta_a^d \delta_c^b - \delta_a^b \delta_c^d
\eea
We then also get
\bea
(\sigma_{(I})_a{}^{[b} (\sigma_{J)})_c{}^{d]} &=& -\delta_{IJ} \delta_{ac}^{bd}\cr
(\sigma_{[I})_a{}^d (\sigma_{J]})_c{}^b &=& \frac{i}{2}\epsilon_{IJK} \((\sigma_K)_a{}^b \delta_c^d - \delta_a^b (\sigma_K)_c{}^d\)
\eea

\section{The fuzzy two-sphere}
The round two-sphere has the isometry group $SO(3)$ which shares the same Lie algebra as $SU(2)$,
\bea
[J_I,J_J] &=& 2i \epsilon_{IJK} J_K
\eea
The Casimir operator is
\bea
J_I J_I &=& N^2-1
\eea
in the dimension-$N$ representation. The fuzzy spherical harmonics are given by
\bea
Y_{\l m}(J) &=& J_{(I_1}\cdots J_{I_\l)}-traces
\eea
Here $m = -\l,...,\l$ runs over $2\l+1$ values. Namely for each fixed length $\l$ there are $2\l+1$ independent spherical harmonics. A symmetric rank-$\l$ tensor where each entry can take $k$ values, has 
\bea
N_{\l,k} &=& \frac{(\l+1)...(\l+k-1)}{1\cdots (k-1)}
\eea
independent components. In our case $k=3$. Removing traces amounts to 
\bea
N_{\l,3} - N_{\l-2,3} = \frac{(\l+1)(\l+2)}{2} - \frac{(\l-1)\l}{2} = 2\l+1
\eea
independent components, which we may label by $m = -\l,...,\l$. If we sum all components of $Y_{\l m}$ for $\l=0,...,N-1$ we get $N^2$ which indicates that we can express any $N\times N$ matrix as an expansion in $Y_{\l m}$ where $\l=1,...,N-1$. If we remove $\l=0$ we get instead $N^2-1$ which corresponds to traceless matrices. All these matrices are hermitian since $J_I$ are hermitian, and any symmetric product of $J_I$ is also hermitian. So by including $\l=0$ we generate all the generators of $U(N)$, and by excluding $\l=0$ we generate all generators of $SU(N)$.

If we rescale 
\bea
\h{J_I} &\cong & \frac{r}{\sqrt{N^2-1}}J_I
\eea
then the correspondence with geometry is provided by the isomorphism
\bea
x_I &\cong & \h{J_I}
\eea
where functions on the sphere are star-multiplied with noncommutativity parameter
\bea
\hbar &=& \frac{2r}{\sqrt{N^2-1}}
\eea
The inner products are related as
\bea
\frac{1}{4\pi r^2}\int_{S^2} d^2 \sigma \sqrt{G} &\cong & \frac{1}{N}\tr_N
\eea
and we may sometimes use the abbreviation
\bea
\int_{S^2} &:=& \frac{1}{4\pi r^2}\int_{S^2} d^2 \sigma \sqrt{G}
\eea

\end{document}